\documentclass[conference]{IEEEtran}
\usepackage{amsmath}
\usepackage{amssymb}
\usepackage{amsfonts}
\usepackage{color}
\usepackage[noadjust]{cite}
\usepackage{graphicx} 
\usepackage[labelsep=period]{caption}
\usepackage[utf8]{inputenc}
\usepackage[english]{babel}
\usepackage[linesnumbered,ruled]{algorithm2e}
\usepackage{subcaption}
\usepackage{cleveref}
\usepackage{tabularx}
\usepackage{geometry}
\newcommand{\argmax}[1]{\underset{#1}{\operatorname{arg}\,\operatorname{max}}\;}

\makeatletter
\newcommand{\multiline}[1]{%
\begin{tabularx}
	{\dimexpr\linewidth-\ALG@thistlm}[t]{@{}X@{}}#1
\end{tabularx}}

\newcommand\fs@betterruled{%
  \def\@fs@cfont{\bfseries}\let\@fs@capt\floatc@ruled
  \def\@fs@pre{\vspace*{5pt}\hrule height.8pt depth0pt \kern2pt}%
  \def\@fs@post{\kern2pt\hrule\relax}%
  \def\@fs@mid{\kern2pt\hrule\kern2pt}%
  \let\@fs@iftopcapt\iftrue}

\makeatother

\begin{document}
\title{IRS-Enhanced OFDM: Power Allocation and Passive Array Optimization
}
\author{\IEEEauthorblockN{Yifei~Yang$^{\star}$$^{\dagger}$, Shuowen~Zhang$^{\star}$, and Rui~Zhang$^{\star}$}
$^{\star}$ Department of Electrical and Computer Engineering, National University of Singapore. \\
			    $^{\dagger}$ Wireline \& Perforating, Halliburton Company.\\
Email: yifeiyang@u.nus.edu,\{elezhsh, elezhang\}@nus.edu.sg
}
\maketitle

\begin{abstract}
Intelligent reflecting surface (IRS) is a promising new technology for achieving spectrum and energy efficient wireless communication systems in the future. By adaptively varying the incident signals' phases/amplitudes and thereby establishing favorable channel responses through a large number of reconfigurable passive reflecting elements, IRS is able to enhance the communication performance of mobile users in its vicinity cost-effectively. In this paper, we study an IRS-enhanced orthogonal frequency division multiplexing (OFDM) system in which an IRS is deployed to assist the communication between a nearby user and its associated base station (BS). We aim to maximize the downlink achievable rate for the user by jointly optimizing the transmit power allocation at the BS and the passive array reflection coefficients at the IRS. Although the formulated problem is non-convex and thus difficult to solve, we propose an efficient algorithm to obtain a high-quality suboptimal solution for it, by alternately optimizing the BS's power allocation and the IRS's passive array coefficients in an iterative manner, along with a customized method for the initialization. Simulation results show that the proposed design significantly improves the OFDM link rate performance as compared to the cases without the IRS or with other heuristic IRS designs.
\end{abstract}

\begin{IEEEkeywords}
Intelligent reflecting surface (IRS), passive array optimization, power allocation, OFDM
\end{IEEEkeywords}
\section{Introduction}
The explosion of mobile data and the ever-increasing demand for higher data rates have continuously driven the advancement of wireless communication technologies in the past decade, such as polar code, massive multiple-input multiple-output (MIMO) and millimeter wave (mmWave) communications, among others. Moreover, a 1000-fold increment in network capacity with ubiquitous connectivity and low latency is envisioned for the forthcoming fifth-generation (5G) wireless network \cite{5g}. Meanwhile, the energy efficiency of future wireless networks is also aimed to be improved by several orders of magnitude so as to maintain the power consumption at increasingly higher data rates. 

Recently, intelligent reflecting surface (IRS) has been proposed as a promising solution to achieve the above goals in a cost-effective way \cite{qqtwc,qqicassp, debtwc,mmwre, metasurmag,qqmag}. Specifically, IRS is a reconfigurable planar array comprising a vast number of passive reflecting elements, which are able to independently induce a phase shift to the incident signal and thus collaboratively alter the reflected signal propagation to achieve desired channel responses in wireless communications. By properly adjusting the phase shifts of IRS's elements, their reflected signals can combine with those from other paths coherently at the receiver to maximize the link achievable rate. Moreover, an amplitude reflection coefficient between zero and one can be designed for the incident signal at each element, so as to further enhance the performance \cite{qqmag}. Hence, different from the conventional half-duplex amplify-and-forward (AF) relay, IRS achieves high beamforming gains by intelligent reflection in a full-duplex manner, thus without consuming any energy or requiring additional time/frequency resource for signal re-generation and re-transmission. It is worth noting that passive reflect-array antennas have been applied in radar and satellite communication systems before. However, their uses in mobile wireless communications are rather limited, since traditional passive arrays only allow for fixed phase-shift patterns once fabricated, and are thus unable to adapt to the dynamic wireless channel due to user mobility. Fortunately, the advances in radio frequency (RF) micro electromechanical systems (MEMS) and metamaterial (e.g., metasurface) have made it feasible to reconfigure the phase shifts in real time \cite{metasur}. Moreover, by varying the resistor load at each element, the reflection amplitude at the IRS can also be flexibly controlled \cite{irsamp}. This thus greatly enhances the functionality and applicability of IRS for wireless communications.

\begin{figure}[t]
    \centering
    \captionsetup{justification=centering}
    \includegraphics[width=0.8\linewidth, keepaspectratio]{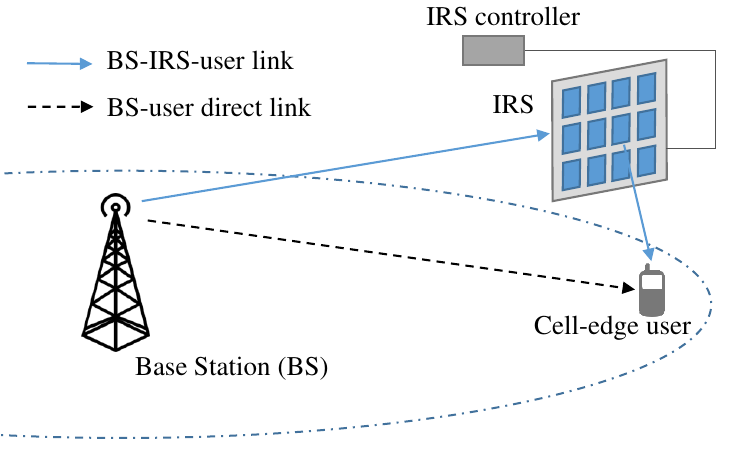}
    \caption{An IRS-enhanced wireless system.}
    \label{fig:irs}
    \vspace{-7mm}
\end{figure}

Prior works on IRS-aided wireless systems \cite{qqtwc,qqicassp, debtwc,mmwre} have considered frequency-flat (non-selective) fading channels for narrow-band communication. However, the design of IRS passive array coefficients for the more general frequency-selective fading channels for broadband communication still remains open, to the authors' best knowledge. Motivated by this, we consider in this paper an IRS-aided orthogonal frequency division multiplexing (OFDM)-based wireless system over frequency-selective channels, as shown in Fig. \ref{fig:irs}. For the purpose of exposition, we consider the case of downlink communication from a base station (BS) to one single user in the vicinity of an IRS, where the IRS and the user are both far away from the BS (e.g., in a cell-edge user scenario). We will extend our results to the uplink communication as well as the general case with multiple users served by the BS/IRS in the journal version of this work. As IRS has a large number of reflecting elements, how to jointly design their reflection coefficients (i.e., phase shifts and amplitude attenuations) so as to achieve the optimal constructive superposition of the reflected signals by the IRS and those from other paths at the user receiver is crucial to maximizing the link achievable rate. However, this is a non-trivial problem to solve under our considered setup due to two main reasons. Firstly, under the frequency-selective channel with multiple paths, the reflection coefficients of the IRS need to cater to the channel gains and delays of all paths to the user, including both the reflected paths by the IRS and the remaining non-reflected paths directly from the BS. Secondly, the achievable rate for the user is determined by both the IRS reflection coefficients and the transmit power allocation at the BS over OFDM subcarriers (SCs), which are intricately coupled and thus need to be jointly optimized. To tackle the above challenges, we formulate a new optimization problem aiming to maximize the downlink achievable rate by jointly optimizing the transmit power allocation at the BS and the passive array coefficients at the IRS, which is non-convex and thus difficult to solve. This motivates us to propose an iterative algorithm to alternately optimize the power allocation and passive array coefficients with a customized initialization scheme, which is guaranteed to converge to at least a locally optimal solution. It is shown by numerical results that the proposed algorithm achieves significantly improved rate performance compared to systems without using the IRS or with heuristic IRS reflection coefficient designs. In addition, it is shown that the proposed initialization scheme is able to achieve a good performance very close to that by the iterative algorithm, and is thus suitable for low-complexity implementation.   

\section{System Model}
We consider a single-user downlink OFDM-based wireless system, wherein an IRS is employed to enhance the communication between a BS and a user, as illustrated in Fig. \ref{fig:irs}. For the purpose of exposition, we assume that the BS and the user are both equipped with a single antenna. The IRS is assumed to comprise $M$ passive reflecting units, denoted by the set $\mathcal{M}=\{1,\dotsc, M\}$, and is connected to a controller, which adjusts the IRS pattern for desired signal reflection. A separate wireless control link serves for information exchange between the IRS controller and the BS on the channel state information (CSI) and other information needed for implementing the joint design of the BS transmission and IRS reflection. It is further assumed that the signals that are reflected by the IRS more than once have negligible power due to severe path loss and are thus ignored. We consider a quasi-static block fading channel model for all channels involved and focus on one particular fading block where the channels remain approximately constant. To obtain the optimal design and the corresponding performance upper bound, we assume that perfect CSI of all channels is available at the BS via channel training and/or feedback from the IRS/user, and thus it can compute the optimal solution and inform it to the IRS for implementation.

Similar to conventional OFDM-based systems, the total bandwidth of the system is equally divided into $N$ orthogonal SCs, denoted by the set $\mathcal{N}=\{0,\dotsc,N-1\}$. Moreover, let $\boldsymbol{p}=[p_0,\dotsc,p_{N-1}]^T\in\mathbb{R}^{N\times 1}$, where each $p_n\geq 0$ denotes the transmit power allocated to SC $n$ at the BS. Assume the total transmission power at the BS is $P$. Thus, the power allocation at the BS should satisfy $\sum_{n=0}^{N-1} p_n\leq P$. 

Let $\boldsymbol{h}_{d}=[\bar{h}_{0},\dotsc,\bar{h}_{L-1},\boldsymbol{0}_{1\times\left(N-L\right)}]^T\in\mathbb{C}^{N\times 1}$ denote the zero-padded $L$-tap baseband equivalent multipath channel of the BS-user direct link, where $\boldsymbol{0}_{a \times b}$ denotes an all-zero matrix of size $a\times b$. Moreover, there exists an $L_0$-tap baseband equivalent multipath channel of the BS-IRS-user link, through which the signal transmitted by the BS is reflected by the IRS before arriving at the receiver of the user. Let $[\boldsymbol{h}_0, \dotsc,\boldsymbol{h}_{L_0-1}, \boldsymbol{0}_{M\times \left(N-L_0\right)}]\in\mathbb{C}^{M\times N}$ denote the zero-padded $L_0$-tap baseband equivalent BS-IRS channel, where each $\boldsymbol{h}_l\in\mathbb{C}^{M\times 1}$ corresponds to the BS-IRS channel at the $l$th tap, $0\leq l\leq L_0-1$. Similarly, let $[\boldsymbol{g}_{0},\dotsc,\boldsymbol{g}_{L_0-1},\boldsymbol{0}_{M\times\left(N-L_0\right)}]^H\in\mathbb{C}^{N\times M}$ denote the baseband equivalent channel of IRS-user link, where each $\boldsymbol{g}_{l}^H\in\mathbb{C}^{1\times M}$ corresponds to the IRS-user channel at the $l$th tap, $0\leq l\leq L_0-1$. At the IRS, each element re-scatters the received signal with an independent reflection coefficient. Specifically, let $\boldsymbol{\phi}=[\phi_1, \dotsc, \phi_M]^T\in\mathbb{C}^{M\times 1}$ denote the IRS reflection coefficients, where each $\phi_m=\beta_m e^{j\theta_m}$ comprises an amplitude coefficient $\beta_m\in[0,1]$ and a phase shift $\theta_m\in[-\pi,\pi)$, i.e., $|\phi_m|\leq 1$. Let $\boldsymbol{\Phi}=\mathrm{diag}\left(\boldsymbol{\phi}\right)$ denote the reflection coefficient matrix of the IRS, where $\boldsymbol{\Phi}\in\mathbb{C}^{M\times M}$ is a square diagonal matrix with the elements of $\boldsymbol{\phi}$ on the main diagonal. The composite BS-IRS-user channel, denoted by $\boldsymbol{h}_{r}\in\mathbb{C}^{N\times 1}$, is thus the concatenation of the BS-IRS channel, IRS reflection, and IRS-user channel, which is given by
\begin{align}
	\boldsymbol{h}_{r}=\begin{bmatrix}
		\boldsymbol{g}_{0}^H\boldsymbol{\Phi}\boldsymbol{h}_{0},
		\dotsc,
		\boldsymbol{g}_{L_0-1}^H\boldsymbol{\Phi}\boldsymbol{h}_{L_0-1},
		\boldsymbol{0}_{1\times\left(N-L_0\right) }
	\end{bmatrix}^T, 
\end{align}
where each $\boldsymbol{g}_{l}^H\boldsymbol{\Phi}\boldsymbol{h}_{l}$ corresponds to the effective BS-IRS-user channel at the $l$th tap, $0\leq l\leq L_0-1$. Hence, the superposed channel impulse response (CIR) from the BS to the user by combining the BS-user (direct) channel and the BS-IRS-user (IRS-reflected) channel is given by
\begin{equation}\label{eqn:hall}
	\tilde{\boldsymbol{h}}=\boldsymbol{h}_{d}+\boldsymbol{h}_{r}.
\end{equation}

Assume OFDM modulation at the BS with a cyclic prefix (CP) of length $\mu$, which is no smaller than $\max(L, L_0)$. Then the channel frequency response (CFR) $\boldsymbol{v}=[v_{0},\dotsc,v_{N-1}]^T\in\mathbb{C}^{N\times 1}$ of the CIR $\tilde{\boldsymbol{h}}$ is given by
\begin{align}
	\boldsymbol{v}=\boldsymbol{F}_N\tilde{\boldsymbol{h}},\label{eqn:vall}
\end{align}
where $\boldsymbol{F}_N\!\in\!\mathbb{C}^{N\times N}$ denotes the discrete Fourier transform (DFT) matrix. Define $\boldsymbol{V}\!=\![\boldsymbol{\nu}_0,\dotsc,\boldsymbol{\nu}_{L_0-1},\boldsymbol{0}_{M\times (N-L_0)}]\!\in\!\mathbb{C}^{M\times N}$, where $\boldsymbol{\nu}_l^H \triangleq\boldsymbol{g}_l^H\mathrm{diag}\left(\boldsymbol{h}_l\right)\in\mathbb{C}^{1\times M}$. We then have $\boldsymbol{\nu}_l^H\boldsymbol{\phi}=\boldsymbol{g}_l^H\boldsymbol{\Phi}\boldsymbol{h}_l$ and $\boldsymbol{h}_{r}=\boldsymbol{V}^H\boldsymbol{\phi}$. Hence, the overall CFR can be rewritten as $\boldsymbol{v}=\boldsymbol{F}_N\tilde{\boldsymbol{h}}=\boldsymbol{F}_N\left(\boldsymbol{h}_{d}+\boldsymbol{V}^H\boldsymbol{\phi}\right)$, and the CFR at each $n$th SC is given by
\begin{equation}\label{eqn:vn}
	v_n=\boldsymbol{f}_n^H\boldsymbol{h}_{d}+\boldsymbol{f}_n^H\boldsymbol{V}^H\boldsymbol{\phi}, \quad n\in\mathcal{N},
	\vspace{-1mm}
\end{equation}
where $\boldsymbol{f}_n^H$ denotes the $n$th row of the DFT matrix $\boldsymbol{F}_N$. The achievable rate in bits per second per Hertz (bps/Hz) is thus obtained as (by accounting for the CP overhead)
\begin{align}
	r(\boldsymbol{p},\boldsymbol{\phi})\!=\!\frac{1}{N\!+\!\mu}\!\sum_{n=0}^{N-1}\! \log_2\!\left(\!1\!+\!\frac{|\boldsymbol{f}_n^H\boldsymbol{h}_{d}\!+\!\boldsymbol{f}_n^H\boldsymbol{V}^H\boldsymbol{\phi}|^2p_n}{\Gamma\sigma^2} \!\right)\!,\label{eqn:rk}
\end{align}
where $\Gamma\geq 1$ is the gap from channel capacity owing to a practical modulation and coding scheme (MCS); the receiver noise is assumed to be independent over all SCs, and is modelled as a circularly symmetric complex Gaussian (CSCG) random variable with mean zero and variance $\sigma^2$. 

\vspace{-1mm}
\section{Problem Formulation}
\vspace{-1mm}
In this paper, we aim to maximize the achievable rate by jointly optimizing the BS transmit power allocation and the IRS reflection coefficients. Therefore, we formulate the following optimization problem 
\begin{align}
\mathrm{(P1)}:\mathop{\mathtt{maximize}}_{\boldsymbol{p},\boldsymbol{\phi}}  &~\sum_{n=0}^{N-1} \log_2\left(1\!+\!\frac{|\boldsymbol{f}_n^H\boldsymbol{h}_{d}\!+\!\boldsymbol{f}_n^H\boldsymbol{V}^H\boldsymbol{\phi}|^2p_n}{\Gamma\sigma^2} \right)	\nonumber\\
	\mathtt{subject\; to}
&~~\sum_{n=0}^{N-1} p_{n}\leq P, \label{eqn:const1}\\
&~~p_n \geq0, \quad \forall n \in \mathcal{N}, \label{eqn:const2}\\
&~~|\phi_m|\leq 1, \quad \forall m\in \mathcal{M}.\label{eqn:const3}
\end{align}
Note that Problem (P1) is a non-convex optimization problem. Particularly, it can be shown that the objective function of (P1) is non-concave over $\boldsymbol{\phi}$; moreover, the variables $\boldsymbol{\phi}$ and $\boldsymbol{p}$ are coupled in the objective function, which makes their joint optimization difficult. To overcome the above challenges, in the following section, we propose an alternating optimization algorithm to find an approximate solution to (P1), by iteratively optimizing one of $\boldsymbol{p}$ and $\boldsymbol{\phi}$ with the other fixed at each time. In addition, we devise a customized method to obtain an initial solution of $\boldsymbol{\phi}$, denoted by $\boldsymbol{\phi}_0$, as the starting point of the proposed alternating optimization algorithm. 

\vspace{-1mm}
\section{Proposed Solution}
\label{sec:sol}
\vspace{-1mm}
\subsection{Power Allocation Optimization Given IRS Coefficients}
\label{sec:pa}
\vspace{-1mm}
Note that given a set of IRS coefficients $\boldsymbol{\phi}$, the CFR $\boldsymbol{v}$ is fixed. The optimal BS transmit power allocation $\boldsymbol{p}$ is thus given by the well-known water-filling (WF) solution \cite{goldsmith}, i.e.,
\begin{equation}\label{eqn:p1}
	p_n=\left(\frac{1}{c_u}-\frac{1}{c_n}\right)^+,\quad\forall n\in\mathcal{N},
	\vspace{-1mm}
\end{equation}
where $\left(x\right)^+\triangleq\max\left(0,x\right)$, $c_n=|v_n|^2/(\Gamma\sigma^2)$ is the effective channel-to-noise power ratio (CNR) for SC $n$, and $c_u$ is the cut-off CNR that satisfies
\begin{align}
	\sum_{n=0}^{N-1} \left(\frac{1}{c_u}-\frac{1}{c_n}\right)^+=P.\label{eqn:p2}
\end{align} 

\subsection{IRS Coefficient Optimization Given Power Allocation}
With given power allocation, Problem (P1) is simplified as 
\begin{align}
\mathrm{(P1.1)}:\mathop{\mathtt{maximize}}_{\boldsymbol{\phi}}  &~\sum_{n=0}^{N-1} \log_2\left(\!1\!+\!\frac{|\boldsymbol{f}_n^H\boldsymbol{h}_{d}\!+\!\boldsymbol{f}_n^H\boldsymbol{V}^H\boldsymbol{\phi}|^2p_n}{\Gamma\sigma^2} \!\right)	\nonumber\\
	\mathtt{subject\; to}
&~~|\phi_m|\leq 1, \quad \forall m\in \mathcal{M}.
\end{align}
It can be shown that (P1.1) is not a convex optimization problem. In the following, we adopt the successive convex approximation (SCA) technique to obtain a locally optimal solution to (P1.1). First, by introducing a set of auxiliary variables $y_n$'s, $a_n$'s, and $b_n$'s, we transform (P1.1) into the following equivalent problem
\begin{align}
\mathrm{(P1.1')}:\mathop{\mathtt{maximize}}_{\boldsymbol{\phi},\{\!y_n\!\},\{\!a_n\!\},\{\!b_n\!\}}  &~\sum_{n=0}^{N-1} \log_2\left(1+\frac{y_np_n}{\Gamma\sigma^2} \right)	\nonumber\\
	\mathtt{subject\; to}
&~|\phi_m|\leq 1, \quad \forall m\in \mathcal{M},\label{eqn:phim}\\
&~a_n\!=\!\Re\{\boldsymbol{f}_n^H\boldsymbol{h}_{d}\!+\!\boldsymbol{f}_n^H\boldsymbol{V}^H\boldsymbol{\phi}\},\;\forall n\!\in\!\mathcal{N},\label{eqn:an}\\
&~b_n\!=\!\Im\{\boldsymbol{f}_n^H\boldsymbol{h}_{d}\!+\!\boldsymbol{f}_n^H\boldsymbol{V}^H\boldsymbol{\phi}\},\;\forall n\!\in\!\mathcal{N},\label{eqn:bn}\\
&~y_n\leq a_n^2+b_n^2, \quad\forall n\in\mathcal{N},\label{eqn:const4}
\vspace{-2mm}
\end{align}
where $\Re({\cdot})$ and $\Im({\cdot})$ denote the real and imaginary part of a complex number, respectively.
Define $\tilde{f}_n(a_n,b_n)\triangleq a_n^2+b_n^2$, which is a convex and differentiable function over $a_n$ and $b_n$. Thus, given any $\tilde{a}_n$ and $\tilde{b}_n$, the first-order approximation of $\tilde{f}_n(a_n,b_n)$ at the point $(\tilde{a}_n,\tilde{b}_n)$ serves as a lower bound to it, i.e.,
\begin{align}
	\tilde{f}_n(a_n,b_n)\!\geq\!\tilde{a}_n^2\!+\!\tilde{b}_n^2\!+\!2\tilde{a}_n(a_n\!-\!\tilde{a}_n\!)\!+\!2\tilde{b}_n(b_n\!-\!\tilde{b}_n\!)\!\triangleq \!f_n(a_n,b_n\!),
\end{align}
where equality holds if and only if $\tilde{a}_n=a_n$ and $\tilde{b}_n=b_n$. Note that $f_n(a_n,b_n)$ is an affine function over $a_n$ and $b_n$, which also has the same gradient over $a_n$ and $b_n$ as $\tilde{f}_n(a_n,b_n)$ at the point $(\tilde{a}_n,\tilde{b}_n)$.

Next, we consider the following optimization problem
\begin{align}
\mathrm{(P1.2)}:\mathop{\mathtt{maximize}}_{\boldsymbol{\phi},\{\!y_n\!\}\!,\{\!a_n\!\}\!,\{\!b_n\!\}\!}  &~\sum_{n=0}^{N-1} \log_2\left(1+\frac{y_np_n}{\Gamma\sigma^2} \right)	\nonumber\\
	\mathtt{subject\; to}
&~(\ref{eqn:phim}),(\ref{eqn:an}),(\ref{eqn:bn})\nonumber\\
&~y_n\leq f_n(a_n,b_n),\quad\forall n\in\mathcal{N}.\label{eqn:const7}
\end{align}
Problem (P1.2) is a convex optimization problem, which can be solved efficiently via existing software in polynomial time with respect to $N$ and $M$, e.g., CVX \cite{cvx}. Therefore, an approximate solution to (P1.1') and thus (P1.1) can be obtained by successively updating $\{\tilde{a}_n\}$ and $\{\tilde{b}_n\}$ based on the optimal solution to (P1.2), which is summarized in Algorithm 1. It can be shown that monotonic convergence of Algorithm 1 is guaranteed, and the obtained solution is a locally optimal solution to (P1.1) \cite{sca}. 

To summarize, the overall iterative algorithm to solve (P1) is given in Algorithm 2. It is worth noting that starting from an initial point denoted by $\boldsymbol{\phi}_0$, the initial value $\tilde{\boldsymbol{\phi}}$ for Algorithm 1 in each iteration of Algorithm 2 is set as the obtained $\boldsymbol{\phi}$ in the previous iteration. It can be shown that the objective value of (P1) is non-decreasing over each iteration of Algorithm 2, which is also upper-bounded by a finite value. Therefore, Algorithm 2 is guaranteed to converge. Moreover, the obtained solution to (P1) can be shown to be at least a locally optimal solution based on \cite{conv}. Note that the performance of Algorithm 2 is critically dependent on the choice of the initial IRS reflection coefficients $\boldsymbol{\phi}_0$. In the following subsection, we propose a customized method for finding $\boldsymbol{\phi}_0$ efficiently. 

\begin{algorithm}[!htb]
\caption{IRS Coefficient Optimization Given Power Allocation via SCA}
\KwIn{$\boldsymbol{h}_d$, $\boldsymbol{V}$, $\boldsymbol{p}$, $\Gamma$, $\sigma^2$, $N$, $M$, $\tilde{\boldsymbol{\phi}}$.}
\KwOut{$\boldsymbol{\phi}$.} 
Set $\tilde{a}_n=\Re\{\boldsymbol{f}_n^H\boldsymbol{h}_{d}+\boldsymbol{f}_n^H\boldsymbol{V}^H\tilde{\boldsymbol{\phi}}\}$, $\tilde{b}_n=\Im\{\boldsymbol{f}_n^H\boldsymbol{h}_{d}+\boldsymbol{f}_n^H\boldsymbol{V}^H\tilde{\boldsymbol{\phi}}\}$, $\forall n\in\mathcal{N}$.

\Repeat{the objective value of (P1.1) with the obtained $\boldsymbol{\phi}$ reaches convergence}
   	{Find the optimal solution of $\{a_n\}$, $\{b_n\}$, and $\boldsymbol{\phi}$ to (P1.2) via CVX with given $\{\tilde{a}_n\}$, $\{\tilde{b}_n\}$, and $\boldsymbol{p}$.
   	
   	$\tilde{a}_n=a_n$, $\tilde{b}_n=b_n$, $\forall n\in\mathcal{N}$.}
\end{algorithm}
\vspace{-7mm}
\begin{algorithm}[!htb]
\caption{Alternating Optimization for Solving (P1)}
\KwIn{$\boldsymbol{h}_d$, $\boldsymbol{V}$, $P$, $\Gamma$, $\sigma^2$, $N$, $M$, $\boldsymbol{\phi}=\boldsymbol{\phi}_0$.}
\KwOut{$\boldsymbol{p}$, $\boldsymbol{\phi}$.}
\Repeat{the objective value of (P1) with the obtained $\boldsymbol{p}$ and $\boldsymbol{\phi}$ reaches convergence}
   {Fixing the IRS coefficients $\boldsymbol{\phi}$, find the WF power allocation $\boldsymbol{p}$ according to (\ref{eqn:vn}), (\ref{eqn:p1}), and (\ref{eqn:p2}).
   
   Fixing the power allocation $\boldsymbol{p}$, given initial $\tilde{\boldsymbol{\phi}}=\boldsymbol{\phi}$, update the IRS coefficients $\boldsymbol{\phi}$ via Algorithm 1.}
\end{algorithm}

\vspace{-6mm}
\subsection{Initialization Method}
\label{sec:init}
\vspace{-1mm}
Note that the IRS is able to increase the link rate mainly due to the increased effective channel power between the BS and the user, by creating an additional strong CIR via the BS-IRS-user channel that can constructively combine with that of the BS-user direct channel. Motivated by this, we propose to design the initial value of $\boldsymbol{\phi}$, i.e., $\boldsymbol{\phi}_0$, by maximizing the effective channel power from the BS to the user, which is given by $\|\tilde{\boldsymbol{h}}\|^2=\left\|\boldsymbol{h}_d+\boldsymbol{V}^H\boldsymbol{\phi}\right\|^2$, with $\|\cdot\|$ denoting the $l_2$ norm. Therefore, we formulate the following optimization problem
\begin{align}
\mathrm{(P2)}:~\mathop{\mathtt{maximize}}_{\boldsymbol{\phi}}  &~\left\|\boldsymbol{h}_d+\boldsymbol{V}^H\boldsymbol{\phi}\right\|^2	\nonumber\\
	\mathtt{subject\; to}
&~~|\phi_m|^2\leq 1, \quad \forall m\in\mathcal{M}.\label{eqn:const5}
\end{align}
Note that Problem (P2) is a non-convex quadratically constrained quadratic problem (QCQP), for which we apply the semidefinite relaxation (SDR) \cite{sdr} technique to obtain an approximate solution for it, as follows. Define $\boldsymbol{A}\triangleq\boldsymbol{V}\boldsymbol{V}^H$ and $\boldsymbol{u}\triangleq\boldsymbol{V}\boldsymbol{h}_d$, Problem (P2) is thus equivalent to 
\begin{align}
	~\mathop{\mathtt{maximize}}_{\boldsymbol{\phi}} &~\boldsymbol{\phi}^H\boldsymbol{A}\boldsymbol{\phi}+\boldsymbol{\phi}^H\boldsymbol{u}+\boldsymbol{u}^H\boldsymbol{\phi} \\
	\mathtt{subject\; to}&~~|\phi_m|^2\leq 1, \quad \forall m\in\mathcal{M}.
\end{align}
Note that $\boldsymbol{\phi}^H\boldsymbol{A}\boldsymbol{\phi}=\mathrm{Tr}\left(\boldsymbol{\phi}^H\boldsymbol{A}\boldsymbol{\phi}\right)=\mathrm{Tr}\left(\boldsymbol{\phi}\boldsymbol{\phi}^H\boldsymbol{A}\right)$; similarly, $\boldsymbol{\phi}^H\boldsymbol{u}=\mathrm{Tr}\left(\boldsymbol{u}\boldsymbol{\phi}^H\right)$ and $\boldsymbol{u}^H\boldsymbol{\phi}=\mathrm{Tr}\left(\boldsymbol{\phi}\boldsymbol{u}^H\right)$ hold, where $\mathrm{Tr}(\cdot)$ denotes the matrix trace. By defining $\boldsymbol{w}=[\boldsymbol{\phi},\boldsymbol{u}]^T$ and $\boldsymbol{W}=\boldsymbol{w}\boldsymbol{w}^H$, we transform (P2) into the following problem
\begin{align}
\mathrm{(P2-SDR)}:~\mathop{\mathtt{maximize}}_{\boldsymbol{W}}  ~&\mathrm{Tr}\left(\boldsymbol{W}\boldsymbol{M}\right)\nonumber\\
	\mathtt{subject\; to}~
&\boldsymbol{W}_{m,m}\leq1, \quad \forall m\in\mathcal{M},\\
&\boldsymbol{W}_{m,m}=|u_{m-M}|, \forall m\!-\!M\!\in\!\mathcal{M},\\
&\boldsymbol{W}\succeq\boldsymbol{0},\label{eqn:const7}
\end{align}
where $\boldsymbol{M}=[\boldsymbol{A},\boldsymbol{I}_M; \boldsymbol{I}_M,\boldsymbol{0}_{M\times M}]$, $\boldsymbol{I}_M$ denotes the identity matrix of size $M\times M$, and the constraint in (\ref{eqn:const7}) ensures $\boldsymbol{W}$ is positive semidefinite. Note that (P2) can be shown to be equivalent to (P2-SDR) with the additional constraint of $\mathrm{rank}(\boldsymbol{W})=1$.  
\begin{algorithm}[b]
\caption{Algorithm for Solving (P2)}

\KwIn{$\boldsymbol{h}_d$, $\boldsymbol{V}$, $M$, $Q$.}
\KwOut{$\boldsymbol{\phi}$.}
Solve (P2-SDR) via CVX and obtain the optimal solution $\boldsymbol{W}^{\star}$. 

Obtain $\boldsymbol{W}_s^{\star}$ by $[\boldsymbol{W}_s^{\star}]_{i,j}=[\boldsymbol{W}^{\star}]_{i,j}$, $i\in\mathcal{M}$, $j\in\mathcal{M}$. 

Compute the EVD of $\boldsymbol{W}_s^{\star}$ by $\boldsymbol{W}_s^{\star}=\boldsymbol{U}\boldsymbol{\Lambda}\boldsymbol{U}^H$.

\eIf{$\mathrm{rank}\left(\boldsymbol{W}^{\star}\right)=1$}
	{$\boldsymbol{\phi}=\boldsymbol{U}\mathrm{Diag}\left(\boldsymbol{\Lambda}^{\frac{1}{2}}\right)$. }
{
\For{$q=1$ \KwTo $Q$}{
   Generate $\tilde{\boldsymbol{r}}^{(q)}\sim\mathcal{CN}\left(\boldsymbol{0},\boldsymbol{I}_{M}\right)$. Obtain $\hat{\boldsymbol{\phi}}^{(q)}=e^{j\arg\left(\boldsymbol{U}\boldsymbol{\Lambda}^{\frac{1}{2}}\tilde{\boldsymbol{r}}^{(q)}\right)}$, where $\arg(\cdot)$ denotes the phase extraction operation. 
   
    Compute the corresponding channel power $P_h^{(q)}=\|\boldsymbol{h}_d+\boldsymbol{V}^H\hat{\boldsymbol{\phi}}^{(q)}\|^2	$.
}

$q^{\star}=\argmax{q=1,\dotsc,Q} P_h^{(q)}$, $\boldsymbol{\phi}=\hat{\boldsymbol{\phi}}^{(q^{\star})}$.
}

\end{algorithm}
Problem (P2-SDR) is a convex semidefinite program (SDP), which can be solved efficiently via existing software, e.g., CVX \cite{cvx}, with polynomial complexity in $M$ \cite{sdr}. Let $\boldsymbol{W}^{\star}$ denote the optimal solution to (P2-SDR). If $\mathrm{rank}\left(\boldsymbol{W}^{\star}\right)\!=\!1$, the relaxation from (P2) to (P2-SDR) is tight and the optimal $\boldsymbol{\phi}$ to Problem (P2) can be obtained as $\boldsymbol{\phi}^{\star}\!=\!\boldsymbol{U}\mathrm{Diag}\left(\boldsymbol{\Lambda}^{\frac{1}{2}}\right)$, where $\boldsymbol{U}\boldsymbol{\Lambda}\boldsymbol{U}^H$ is the eigenvalue decomposition (EVD) of the upper left $M\!\times\! M$ submatrix of $\boldsymbol{W}^{\star}$, denoted by $\boldsymbol{W}_s^{\star}$, and $\mathrm{Diag}(\boldsymbol{\Lambda}^{\frac{1}{2}})$ denotes the column vector formed by the main diagonals of $\boldsymbol{\Lambda}^{\frac{1}{2}}$. On the other hand, if $\mathrm{rank}\left(\boldsymbol{W}^{\star}\right)\!>\!1$, the optimal objective value of Problem (P2-SDR) serves as an upper bound to that of Problem (P2) and additional processing is required to construct a rank-one solution according to $\boldsymbol{W}^{\star}$. In particular, we consider a customized Gaussian randomization method \cite{sdrrand} to find an approximate solution to Problem (P2). To enhance the performance of the proposed algorithm, a number (denoted by $Q$) of $\hat{\boldsymbol{\phi}}$'s are generated based on $\boldsymbol{W}^{\star}$, from which the one that yields the largest objective value of Problem (P2) is selected. The overall algorithm for solving (P2) is summarized in Algorithm 3, where the output $\boldsymbol{\phi}$ of Algorithm 3 is then set as the initial $\boldsymbol{\phi}_0$ for Algorithm 2.

\vspace{-1mm}
\section{Numerical Results}
\vspace{-1mm}
In this section, we examine the performance of our proposed algorithm via numerical results. We consider $N=64$ and $L=L_0=16$ for both the BS-user link (i.e., the direct link) and the BS-IRS-user link (i.e., the reflected link), among which $8$ taps at random delays are non-zero for each link and are modeled as CSCG random variables with an exponential power delay profile. The CP length is set as $\mu=16$. The total average channel power of the reflected link over all taps is defined as $P_r=\sum_{l=0}^{L_0-1}\mathbb{E}[\|\boldsymbol{g}_l\|^2\|\boldsymbol{h}_l\|^2]$, and that of the direct link is given by $P_d=\mathbb{E}[\|\boldsymbol{h}_d\|^2]$. The average signal-to-noise ratio (SNR) is thus given by $\gamma=P(P_r+P_d)/(N\sigma^2)$, while we normalize the total average channel power of the two links as $P_d+P_r=1$ for convenience, unless stated otherwise. Let $\alpha=P_r/P_d$ denote the average power ratio of the reflected link to the direct link. Hence, $\alpha\rightarrow 0$ indicates that the user is located far away from the IRS, thus its channel with the BS is dominated by the BS-user direct link; while on the other hand, $\alpha\rightarrow\infty$ indicates that the user is located in close vicinity of the IRS. The SNR gap is set as $\Gamma=8.8$ dB, and the number of randomizations in Algorithm 3 is chosen as $Q=50$. All the results are averaged over 100 independent channel realizations.

For comparison, we consider the following benchmark schemes:
\begin{enumerate}
	\item \textbf{Channel Power Maximization (CPM)}: In this scheme, we adopt the IRS coefficients as $\boldsymbol{\phi}_0$ obtained via the initialization method based on CPM proposed in Section \ref{sec:init}, and the WF-based power allocation based on $\boldsymbol{\phi}_0$.
	\item \textbf{Random Phase}: We assume the IRS coefficients have random phase and maximum amplitude. As the channel coefficients are randomly generated, this is equivalent to setting $\boldsymbol{\phi}=\boldsymbol{1}$, based on which we obtain the WF-based power allocation. Note that the IRS behaves like a lossless reflective mirror in this case.
	\item \textbf{Without IRS}: We consider the WF power allocation and achievable rate based on the BS-user direct link only.
\end{enumerate}

\begin{figure}[t]
    \centering
    \captionsetup{justification=centering}
    \vspace{-2mm}
    \includegraphics[width=0.82\linewidth, keepaspectratio]{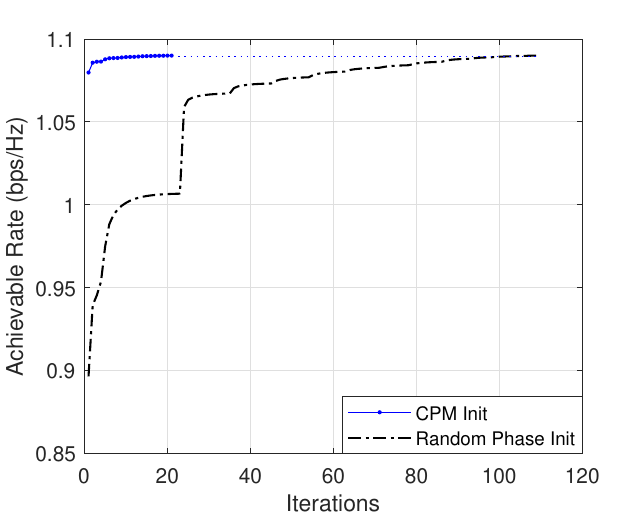}
    \vspace{-1mm}
    \caption{Convergence behavior of Algorithm 2.}
    \label{fig:ini_comp}
    \vspace{-5mm}
\end{figure}
First, we evaluate the convergence behavior of Algorithm 2. The number of reflecting elements is set as $M=20$, and the power ratio of the reflected link to the direct link is set as $\alpha=10$. For comparison with the CPM-based initialization method proposed in Section \ref{sec:init}, we consider a benchmark initialization method with random phase (or $\boldsymbol{\phi}_0=\boldsymbol{1}$). Fig. \ref{fig:ini_comp} shows the achievable rate over iterations at SNR $=15$ dB for a random channel realization. Monotonic convergence is observed for both initialization methods, which is consistent with our discussions in Section \ref{sec:sol}. Moreover, it is observed that the proposed CPM-based method converges much faster compared to the random phase method (i.e., $21$ versus $109$ iterations), while both methods achieve the same converged achievable rate (i.e., $1.0900$ bps/Hz). This thus validates the efficiency of the proposed CPM-based initialization method. 

\begin{figure}[t]
    \centering
    \captionsetup{justification=centering}
    \includegraphics[width=0.82\linewidth, keepaspectratio]{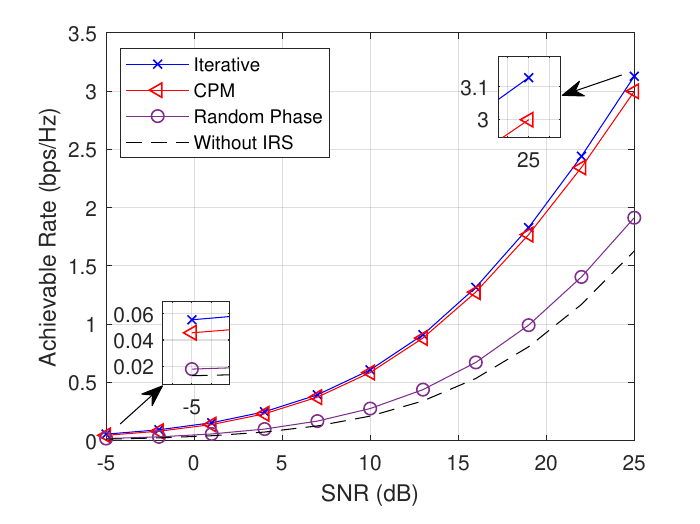}
    \vspace{-1mm}
    \caption{Achievable rate versus SNR.}
    \label{fig:su_snr}
    \vspace{-6mm}
\end{figure}
Fig. \ref{fig:su_snr} shows the performance of the iterative algorithm and the benchmark schemes at different SNR values, with $M\!=\!20$ and $\alpha\!=\!10$. It is observed that all the three schemes with IRS outperform the scheme without IRS, due to the IRS-enhanced average channel power from the BS to the user. Moreover, the proposed iterative algorithm and CPM-based initialization scheme both achieve significantly improved achievable rates over the random phase scheme, since the direct channel and the reflected channel are superposed more constructively via designing the IRS reflection coefficients. Furthermore, it is observed that the performance of the proposed CPM-based initialization scheme is very close to that of the iterative algorithm, and the performance gap (in terms of percentage increment) decreases as the SNR increases. Therefore, this scheme is suitable for practical implementation with lower complexity.   

\begin{figure}[t]
    \vspace{-2mm}    
    \centering
    \captionsetup{justification=centering}
    \includegraphics[width=0.82\linewidth, keepaspectratio]{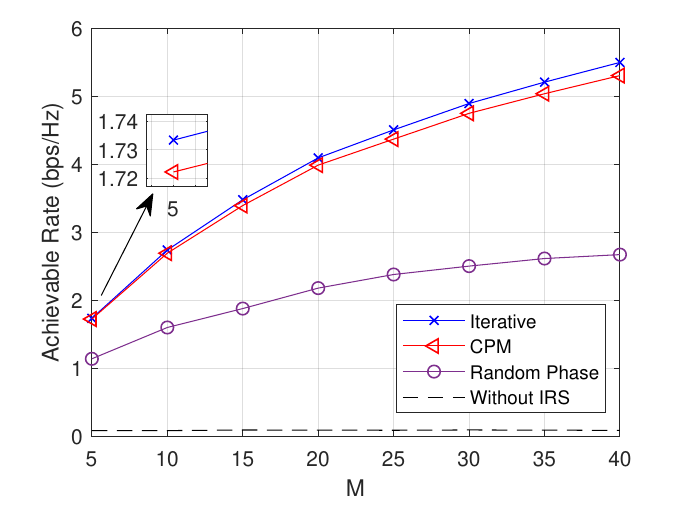}
    \vspace{-1mm}
    \caption{Achievable rate versus number of reflecting elements.}
    \label{fig:su_m}
    \vspace{-5mm}
\end{figure}

Fig. \ref{fig:su_m} compares the performance of the iterative algorithm and the benchmark schemes versus $M$, where we set the reference SNR with $M=1$ as $\bar{\gamma}=5$ dB and the reference power ratio with $M=1$ as $\bar{\alpha}=10$. It is observed that the achievable rates for both the proposed iterative algorithm and initialization scheme increase with $M$, owing to the passive beamforming gain harvested by properly designing the IRS reflection coefficients according to the CSI; while on the other hand, the achievable rate for the random phase scheme increases at a much slower rate with $M$. Moreover, it is observed that the performance gain of the proposed schemes over the scheme without IRS or with random phase becomes more pronounced as $M$ increases.

\begin{figure}[t]
    \centering
    \captionsetup{justification=centering}
    \includegraphics[width=0.82\linewidth, keepaspectratio]{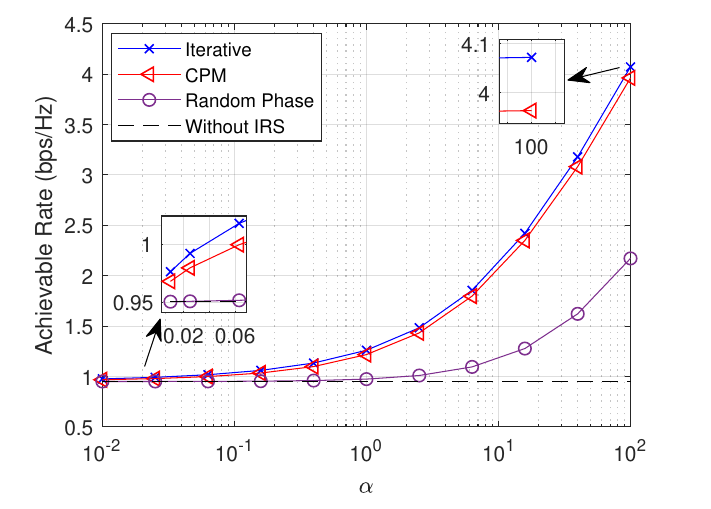}
    \vspace{-1mm}
    \caption{Achievable rate versus reflected to direct link power ratio.}
    \label{fig:su_ratio}
    \vspace{-7mm}
\end{figure}

Fig. \ref{fig:su_ratio} shows the performance of the iterative algorithm and the benchmark schemes versus the reflected to direct link power ratio $\alpha$ with $M=20$. For ease of illustration, we set the reference SNR with only the direct link as $\gamma_d=PP_d/(N\sigma^2)=10$ dB, thus the overall SNR is $\gamma=(1+\alpha)\gamma_d$. Note that as $\alpha$ increases, this corresponds to a practical scenario where the user maintains a fixed distance with the BS (e.g., on a circle centered at the BS), and gradually moves towards the IRS. It is observed that when $\alpha$ is very small, all the schemes yield similar performance since the user is far away from the IRS, whose effect is thus negligible. On the other hand, as $\alpha$ increases, the performance gain of the proposed iterative algorithm and CPM-based initialization method over the benchmark schemes increase drastically, due to the increased dominance of the IRS reflected link over the direct link. Moreover, it is observed that all the schemes with IRS achieve significantly improved performance compared to that without IRS when $\alpha$ is large. This indicates that even though the user is located far away from the BS in the cell-edge scenario, a nearby IRS is effective in enhancing the link rate.
\vspace{-2mm}
\section{Conclusion}
In this paper, we proposed a novel approach to enhance the achievable rate of an OFDM system by utilizing the IRS, and jointly designed the IRS reflection coefficients and BS transmit power allocation to maximize the link achievable rate. By leveraging optimization techniques, we proposed computationally efficient methods to find high-quality suboptimal solutions for the formulated problem. Numerical results showed the effectiveness of IRS in boosting the achievable rate of a cell-edge user aided by the IRS. The proposed initialization method was also shown to achieve very close rate performance compared to the proposed iterative method, but with much lower complexity for implementation. 

\bibliographystyle{IEEEtran}
\bibliography{irs}

\begin{thebibliography}{10}
\providecommand{\url}[1]{#1}
\csname url@samestyle\endcsname
\providecommand{\newblock}{\relax}
\providecommand{\bibinfo}[2]{#2}
\providecommand{\BIBentrySTDinterwordspacing}{\spaceskip=0pt\relax}
\providecommand{\BIBentryALTinterwordstretchfactor}{4}
\providecommand{\BIBentryALTinterwordspacing}{\spaceskip=\fontdimen2\font plus
\BIBentryALTinterwordstretchfactor\fontdimen3\font minus
  \fontdimen4\font\relax}
\providecommand{\BIBforeignlanguage}[2]{{%
\expandafter\ifx\csname l@#1\endcsname\relax
\typeout{** WARNING: IEEEtran.bst: No hyphenation pattern has been}%
\typeout{** loaded for the language `#1'. Using the pattern for}%
\typeout{** the default language instead.}%
\else
\language=\csname l@#1\endcsname
\fi
#2}}
\providecommand{\BIBdecl}{\relax}
\BIBdecl

\bibitem{5g}
J.~G. Andrews, S.~Buzzi, W.~Choi, S.~Hanly, A.~Lozano, A.~C.~K. Soong, and
  J.~C. Zhang, ``What will \protect{5G} be?'' \emph{IEEE J. Sel. Commun.},
  vol.~32, no.~6, pp. 1065--1082, Jun. 2014.

\bibitem{qqtwc}
Q.~Wu and R.~Zhang, ``Intelligent reflecting surface enhanced wireless network
  via joint active and passive beamforming,'' to appear in \textit{IEEE Trans.
  Wireless Commun.}, 2019.

\bibitem{qqicassp}
------, ``Beamforming optimization for intelligent reflecting surface with
  discrete phase shifts,'' in \emph{Proc. IEEE Int. Conf. Acoustics. Speech.
  Signal Process.(ICASSP)}, Brighton, UK, May 2019, pp. 7830--7833.

\bibitem{debtwc}
C.~Huang, A.~Zappone, G.~C. Alexandropoulos, M.~Debbah, and C.~Yuen,
  ``Reconfigurable intelligent surfaces for energy efficiency in wireless
  communication,'' to appear in \textit{IEEE Trans. Wireless Commun.}, 2019.

\bibitem{mmwre}
X.~Tan, Z.~Sun, D.~Koutsonikolas, and J.~M. Jornet, ``Enabling indoor mobile
  millimeter-wave networks based on smart reflect-arrays,'' in \emph{Proc. IEEE
  Conf. Comput. Commun. (INFOCOM)}, Honolulu, HI, Apr. 2018, pp. 270--278.

\bibitem{metasurmag}
C.~Liaskos, S.~Nie, A.~Tsioliaridou, A.~Pitsillides, S.~Ioannidis, and
  I.~Akyildiz, ``A new wireless communication paradigm through
  software-controlled metasurfaces,'' \emph{IEEE Commun. Mag.}, vol.~56, no.~9,
  pp. 162--169, Sep. 2018.

\bibitem{qqmag}
Q.~Wu and R.~Zhang, ``Towards smart and reconfigurable environment: Intelligent
  reflecting surface aided wireless network,'' to appear in \textit{IEEE
  Commun. Mag.}, 2019.

\bibitem{metasur}
T.~J. Cui, M.~Q. Qi, X.~Wan, J.~Zhao, and Q.~Cheng, ``Coding metamaterials,
  digital metamaterials and programmable metamaterials,'' \emph{Light: Science
  \& Applications}, vol. 3, e218, Oct. 2014.

\bibitem{irsamp}
H.~Yang \emph{et~al.}, ``Design of resistor-loaded reflectarray elements for
  both amplitude and phase control,'' \emph{IEEE Antennas Wireless Propag.
  Lett.}, vol.~16, pp. 1159--1162, 2017.

\bibitem{goldsmith}
A.~Goldsmith, \emph{Wireless Communications}.\hskip 1em plus 0.5em minus
  0.4em\relax Cambridge University Press, 2005.

\bibitem{cvx}
M.~Grant and S.~Boyd, ``\protect{CVX: Matlab Software for Disciplined Convex
  Programming},'' \protect{Version} 2.1, Dec. 2018 [Online]. Available:
  http://cvxr.com/cvx/.

\bibitem{sca}
B.~R. Marks and G.~P. Wright, ``A general inner approximation algorithm for
  nonconvex mathematical programs,'' \emph{Operations Research}, vol.~26,
  no.~4, pp. 681--683, Jul. 1978.

\bibitem{conv}
M.~{Hong}, M.~{Razaviyayn}, Z.~{Luo}, and J.~{Pang}, ``A unified algorithmic
  framework for block-structured optimization involving big data: With
  applications in machine learning and signal processing,'' \emph{IEEE Signal
  Process. Mag.}, vol.~33, no.~1, pp. 57--77, Jan. 2016.

\bibitem{sdr}
Z.-Q. Luo, W.-K. Ma, A.~M.-C. So, Y.~Ye, and S.~Zhang, ``Semidefinite
  relaxation of quadratic optimization problems,'' \emph{IEEE Signal Process.
  Mag.}, vol.~27, no.~3, pp. 20--34, May 2010.

\bibitem{sdrrand}
S.~Zhang, R.~Zhang, and T.~J. Lim, ``Constant envelope precoding for
  \protect{MIMO} systems,'' \emph{IEEE Trans. Commun.}, vol.~66, no.~1, pp.
  149--162, Jan. 2018.

\end{thebibliography}
\end{document}